\documentclass[conference, 10pt]{IEEEtran}

\usepackage{cite}
\usepackage{url}
\usepackage{graphicx}
\usepackage{color}
\usepackage{placeins}
\usepackage{float}
\usepackage{tabularx,colortbl}
\usepackage{ifthen}

\usepackage{pgf,tikz}
\usetikzlibrary{positioning}
\usetikzlibrary{arrows.meta}
\usepackage{pgfplotstable}
\usepackage{pgfplots}

\usepackage{amsmath,amssymb,amsfonts}
\usepackage{algorithmic}
\usepackage{textcomp}
\usepackage{xcolor}
\usepackage{mathtools}
\usepackage{todonotes}
\usepackage{siunitx}
\usepackage[belowskip=-15pt,aboveskip=0pt]{caption}

\newcommand{\vt}[1]{\boldsymbol{#1}} 

\newcommand{\veg}[1]{\bm{#1}}               
\newcommand{\mat}[1]{\mathbf{#1}}       
\newcommand{\matd}[1]{\mathbb{#1}}   
\newcommand{\uv}[1]{\hat{\veg{#1}}}         
\newcommand{\op}[1]{\mathcal{#1}}           

\newcommand{\dd}{\mathrm{d}}  

\newcommand{\e}{\mathrm{e}}

\usepackage[utf8]{inputenc}

\usepackage[T1]{fontenc}
\usepackage{subcaption}
\usepackage{float}
\usepackage{tikz}
\usepackage{pgfplots}
\usepackage{pgfplotstable}
\usepackage{siunitx}
\usepackage{cleveref}
\pgfplotsset{compat=newest}

\usepackage{amsthm}
\DeclareMathAlphabet{\mathbfsf}{\encodingdefault}{\sfdefault}{bx}{n}
\usepackage{bm}
\usepackage[OMLmathsfit]{isomath}

\makeatletter

\newcounter{author}
\renewcommand{\author}[2][]{
   \stepcounter{author}
   \@namedef{author@\theauthor}{#2}
   \@namedef{authorlabel@\theauthor}{#1}
}

\newcounter{address}
\newcommand{\address}[2][]{
   \stepcounter{address}
   \@namedef{address@\theaddress}{#2}
   \@namedef{addresslabel@\theaddress}{#1}
}

\newcommand{\alsep}{and}

\def\newmaketitle{\par%
  \begingroup%
  \normalfont%
  \def\thefootnote{}
  \def\footnotemark{}
  \let\@makefnmark\relax
  \footnotesize
  \footnotesep 0.7\baselineskip
  \normalsize%
  \twocolumn[\thenewmaketitle\@IEEEaftertitletext]%
  \if@IEEEusingpubid
     \enlargethispage{-\@IEEEpubidpullup}%
  \fi
  \endgroup
  \setcounter{footnote}{0}\let\maketitle\relax\let\@maketitle\relax
  \gdef\@thanks{}%
  \let\thanks\relax}

\def\thenewmaketitle{
  \newpage
  \begin{center}%
    \vskip0.2em{\Huge\@IEEEcompsoconly{\sffamily}\@IEEEcompsocconfonly{\normalfont\normalsize\vskip 2\@IEEEnormalsizeunitybaselineskip
   \bfseries\large}\@title\par}\vskip1.0em\par%
    \vspace{1ex}
    \newcounter{c@author}
    \newcounter{c@tmp}
    \ifthenelse{\value{author}=2}{%
      \newcommand{\liand}{ and }}{%
      \newcommand{\liand}{, and }}
    \ifthenelse{\value{address}<2}{%
      \@nameuse{author@1}%
      \stepcounter{c@author}%
      \whiledo{\value{c@author}<\value{author}}{%
        \setcounter{c@tmp}{\value{author}}%
        \addtocounter{c@tmp}{-\value{c@author}}%
        \ifthenelse{\value{c@tmp}=1}{%
          \renewcommand{\alsep}{\liand}}{\renewcommand{\alsep}{, }}%
        \stepcounter{c@author}\alsep \@nameuse{author@\thec@author}}\\%
    }
    {
      \@nameuse{author@1}${}^{(\ref{\@nameuse{authorlabel@1}})}$%
      \stepcounter{c@author}%
      \whiledo{\value{c@author}<\value{author}}{%
      \setcounter{c@tmp}{\value{author}}%
      \addtocounter{c@tmp}{-\value{c@author}}%
      \ifthenelse{\value{c@tmp}=1}{%
        \renewcommand{\alsep}{\liand}}{\renewcommand{\alsep}{, }}%
      \stepcounter{c@author}\alsep \@nameuse{author@\thec@author}%
        ${}^{(\ref{\@nameuse{authorlabel@\thec@author}})}$%
      }
    }
    \vspace{0.2ex}

    \ifthenelse{\value{address}>0}{%
      \ifthenelse{\value{address}=1}{
        {\@nameuse{address@1}}
      }
      {
        \newcounter{c@address}

        \begin{center}
        \whiledo{\value{c@address}<\value{address}}
        {
          \refstepcounter{c@address}
            ${}^{(\thec@address)}$\,%
              \label{\@nameuse{addresslabel@\thec@address}}%
              \@nameuse{address@\thec@address}\\ %
        }
        \end{center}
      } 
    }
    {
      \relax
    }
  \end{center}
}

\makeatother

\title{Laplacian Filters for Integral Equations: Further Developments and Fast Algorithms}

\author[imt]{Adrien Merlini}
\author[imt]{Clément Henry}
\author[polito]{Davide Consoli}
\author[polito]{Lyes Rahmouni}
\author[polito]{Francesco P. Andriulli}

\address[polito]{Politecnico di Torino, Turin, Italy}
\address[imt]{IMT Atlantique, Brest, France}

\begin{document}

\newmaketitle
\begin{abstract}
This paper extends the concept of Laplacian filtered quasi-Helmholtz decompositions we have recently introduced, to the basis-free projector-based setting. This extension allows the discrete analyses of electromagnetic integral operators spectra without passing via an explicit Loop-Star decomposition as previously done. We also present a fast scheme for the evaluation of the filters in quasi linear complexity in the total number of unknowns. Together with the fact that only a logarithmic number of these filters are required for solving the h-refinement breakdown of electric field integral equation, this results in an effective preconditioner that rivals Calderón strategies in performance without relying on barycentric refinements. Numerical results confirm the theoretically predicted behavior and the effectiveness of the approach.
\vspace{-0.5cm}
\end{abstract}

\section{Introduction}
The concept of Laplacian filtered Loop and Star bases was recently introduced \cite{8879168} to obtain a discrete block spectral decomposition of the Laplacian on a general geometry for preconditioning the electric field integral equation (EFIE). Although the approach could be used to obtain a full graph basis with a detailed spectral resolution, it was shown that a logarithmic number of spectral filtered subdivisions was sufficient to regularize the otherwise ill-conditioned operator as also suggested by standard wavelet preconditioning theory (see \cite{9580445} and references therein). An alternative to standard Loop and Star bases is a quasi-Helmholtz decomposition approach based on projectors \cite{9580445}. Following the same paradigm in this work we perform the transition from the Laplacian filtered Loop and Star bases in \cite{8879168} to projector Laplacian filters that can effectively perform spectral analyses of the EFIE and of related operators and thus produce well conditioned and rapidly converging formulations, among other things. In addition, we complement the contribution by presenting a fast approach that allows for the application of each projector in quasi-linear complexity. Theoretical considerations will be complemented by numerical results to show the practical impact of the proposed strategies.

\section{Background and Notation}
Consider a Lipschitz boundary $\Gamma$ modeling a perfectly electrically conducting (PEC) object with outward pointing normal $\uv{n}$. Solving the EFIE yields the electric current surface density $\vt{J}$ induced by a time harmonic incident wave $\vt{E}^i$. The EFIE reads
\vspace{-0.2cm}
\begin{equation}
 \label{EFIE}
  	\op{T} \vt{J}=\op{T}_s \vt{J} +  \op{T}_h \vt{J}=-\uv{n}(\vt{r}) \times \vt{E^i}
  \end{equation}
where  $\op{T}_s \vt{J} = \uv{n}(\vt{r}) \times  i k  \int_\Gamma \frac{\e^{i k\| \vt{r} - \vt{r'} \|}}{4\pi \|\vt{r} - \vt{r'} \|}   \vt{J}(\vt{r'})  \dd S(\vt{r'}) $ and $\op{T}_h \vt{J} = -\uv{n}(\vt{r}) \times \frac{1}{i k} \nabla_{\vt{r}} \int_\Gamma \frac{\e^{i k\| \vt{r} - \vt{r'} \|}}{4\pi\|\vt{r} - \vt{r'} \|}  \nabla_{\vt{r'}} \cdot \vt{J}(\vt{r'})  \dd S(\vt{r'}) $ with $k$  the wavenumber. The EFIE is discretized in a standard way by approximating  $\Gamma$ with a mesh of triangular elements of average edge length $h$ and expanding the current as $ \vt{J} = \sum\nolimits_{n = 1}^N {j_n } \vt{f}_n (\vt{r}) $, where $\vt{f}_n (\vt{r})$ are the Rao-Wilton-Glisson (RWG) basis functions. Finally, testing the equation with $\uv{n}(\vt{r}) \times \vt{f}_n (\vt{r})$ functions, the linear system $\mat{T} \mat{j} =\mat{e}$ is obtained. This system is lamentably very ill-conditioned for  both decreasing frequencies and increasing mesh densities with $\text{cond} (\mat{T}) \lesssim 1/(hk)^2$\cite{9580445}.
In the following we will use the normalized loop-to-RWG and star-to-RWG transformation matrices $\mat{\Lambda}$ and $\mat{\Sigma}$ whose explicit definition is omitted due to space limitations but can be found in \cite{9580445}. The normalization is chosen so that $\mat\Lambda^T\mat\Lambda$ and $\mat\Sigma^T\mat\Sigma$ are respectively the vertices- and the cells- based graph Laplacians \cite{9580445}.
\section{Laplacian Filters: the projector approach}
Laplacian filters were exploited in  \cite{8879168} to render the EFIE matrix spectrally block diagonal to precondition it in a ``wavelet-like manner''. In doing so we obtained a filtered Loop-Star decomposition based on filtered basis functions. That approach however, did not have the versatility of basis-free quasi Helmholtz projectors that is especially useful when dealing with complex geometries. To overcome this limitation, we adapt in this work our previously proposed filters to the projector's framework. The general form of the projector filters we propose, a definition that allow for normalization and geometry matrices when needed, reads
\begin{align}
    \mat P_\epsilon^\Lambda &=\sqrt{\mat G} \mat\Lambda \mat A\left(\mat A^T\mat\Lambda^T\mat G\mat\Lambda\mat A\right)^{+}_\epsilon\mat A^T \mat\Lambda^T\sqrt{\mat G}\,, \\
    \mat P_\epsilon^\Sigma &=\sqrt{\mat G^{-1}} \mat\Sigma \mat B\left(\mat B^T\mat\Sigma^T\mat G^{-1}\mat\Sigma\mat B\right)^{+}_\epsilon \mat B^T\mat\Sigma^T\sqrt{\mat G^{-1}}\,,
\end{align}
where $+$ denotes pseudoinversion and  $\left(\mat A^T\mat\Lambda^T\mat G\mat\Lambda\mat A\right)^{+}_\epsilon$ and $\left(\mat B^T\mat\Sigma^T\mat G^{-1}\mat\Sigma\mat B\right)^{+}_\epsilon$ are the filtered Primal and Dual Laplacian theoretically obtained by taking the SVD of each matrix and cutting the singular values at a relative error $\epsilon$ with respect to the first one. In practice, because of the cost of the SVD, those matrices should be obtained with other strategies. In \cite{8879168} we proposed a Butterworth-like filtering approach which is effective when the filtering spectral point is fixed. Here we will present a different strategy that will work without this constraint. Note that the matrices  $\mat G$, $\mat A$, and $\mat B$ are not always required, depending on the formulation and can be set to an identity in case of need. Finally, the filters defined above are those applicable to the normalized operators (i.e. operators discretized with orthonormal bases) and thus the square root matrices appearing in the definitions will simplify or will be squared with the normalization matrices of the operators to be preconditioned. In other words the computation of these matrix square roots will not be necessary.
For the sake of completeness we also define the dual filters that can be used for operators discretized with dual functions (such as the Buffa-Christiansen (BC) or the Chen-Wilton (CW) functions).
\begin{align}
    \matd P_\epsilon^\Lambda &=\sqrt{\mat G_d^{-1}} \mat\Lambda \mat C\left(\mat A^T\mat\Lambda^T\mat G_{d}^{-1}\mat\Lambda\mat C\right)^{+}_\epsilon\mat A^T \mat\Lambda^T\sqrt{\mat G_d^{-1}}\,, \\
    \matd P_\epsilon^\Sigma &=\sqrt{\mat G_d} \mat\Sigma \mat D\left(\mat B^T\mat\Sigma^T\mat G_{d}\mat\Sigma\mat D\right)^{+}_\epsilon \mat B^T\mat\Sigma^T\sqrt{\mat G_d}\,.
\end{align}
On simply connected geometries the complementarity properties $\mat P_0^\Lambda=\mat I-\mat P_0^\Sigma$ and $\matd P_0^\Lambda=\mat I-\matd P_0^\Sigma$ can be proven. On non-simply connected geometries the filters above need to be complemented with the harmonic projectors $\mat P^H =\mat I-\mat P_0^\Sigma-\mat P_0^\Lambda$ and $\matd P^H =\mat I-\matd P_0^\Sigma-\matd P_0^\Lambda$ for standard and dual meshes respectively.
\begin{figure} \centering
 \input{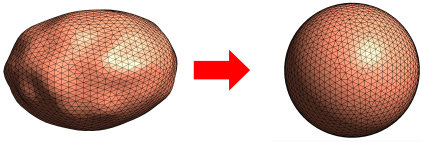}
 \caption{Mesh to Sphere Mesh Morphing}
 \label{fig:morphing}
\end{figure}
\section{Fast Filtering}
Consider the case of an homogeneous mesh so that all matrices $\mat G$, $\mat G_d$, $\mat A$, $\mat B$, $\mat C$,$\mat D$ are identities to focus on filtering the graph Laplacians $\mat\Lambda^T\mat\Lambda$ and $\mat\Sigma^T\mat\Sigma$. This assumption is only introduced to simplify the treatment, but can be lifted by leveraging spectral equivalences and  bandlimitedness of the geometry.  We also assume the structure to be simply connected. The algorithm below will never build the filtered Laplacian matrix itself, but it will perform  a spectrally equivalent matrix vector multiplication of the filter with the vector to be analyzed, in quasi linear complexity. To do so, we followed for these matrices the philosophy of a morphing approach we have proposed previously in
\cite{9539728}. The algorithm is the following:
    1) Map the geometry and its mesh onto the unitary sphere and its induced mesh (see Fig.~\ref{fig:morphing}, note that the graph Laplacians do not change in this case).
    2) Evaluate the projectors by using as $\mat G$ the RWG (or BC in the dual case) matrix of the obtained sphere after mapping. This produces from the graph Laplacian a variational Laplacian corresponding to a Galerkin discretization with scalar linear Lagrange interpolants (dual piecewise-linear interpolants in the dual case).
    3) Obtain the multiplication by leveraging a fast spherical filter: this is obtained by interpolating the pyramidal potentials in the fast spherical filter grid, multiplying times the Laplacian spherical harmonics transformed diagonal with proper zeroing of the spectrum depending on the filter to obtain and, finally, interpolating to the mesh to retrieve the final result. This step it is quite similar to what it is done in high frequency fast multipole method and can be achieved in quasi-linear complexity.
    \begin{figure} 
\centering
\includegraphics[width=0.43\textwidth]{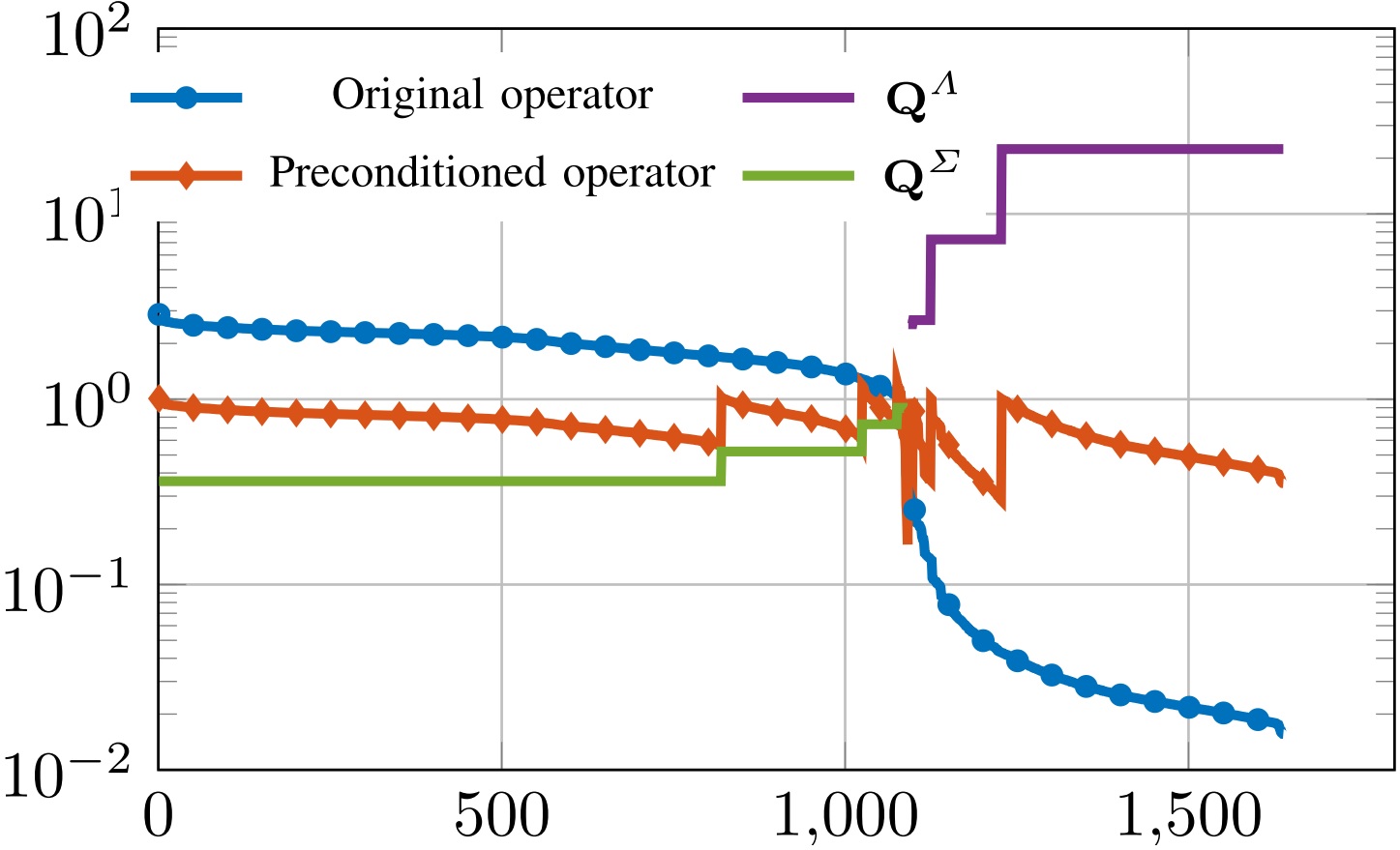}
 \caption{Effect of the wavelet-like scaled projectors}
 \label{fig:spectrum}
\end{figure}
\section{An application: a wavelet-style preconditioner}
As also explained in \cite{8879168}, our Laplacian filtering approach can have several applications and one of them is certainly in preconditioning since, differently from quasi-Helmholtz projectors that can solve the low-frequency breakdown, but not the dense-mesh breakdown, Laplacian filters can solve both using a ``wavelet philosophy'' requiring only a number of filters scaling logarithmically with the number of unknowns. Practically, by defining the projectors
$\mat W^{\Lambda,\Sigma}_j=\mat P^{\Lambda,\Sigma}_{2^{-j}}-\mat P^{\Lambda,\Sigma}_{2^{-j+1}}$ with $j=1,\ldots,L$, $L=\min (\log N_{\Lambda},\log N_{\Sigma})$, and $\mat W^{\Lambda,\Sigma}_0=\mat P^{\Lambda,\Sigma}_{1}$.
a preconditioner of additive Schwarz kind for the EFIE can then be built as $\mat Q\mat T\mat Q$ with
\begin{equation}\label{eq:6}
   \mat Q= \sum_{j=0}^{L} \frac{\mat W^{\Sigma}_j }{\sqrt{\|\mat W^{\Sigma}_j \mat T \mat W^{\Sigma}_j)\|}} +\sum_{j=0}^{L}\frac{ \mat W^{\Lambda}_j }{\sqrt{\|\mat W^{\Lambda}_j \mat T \mat W^{\Lambda}_j)\|}}\,.
\end{equation}
 To allow for a better understanding of the spectral analysis capabilities of the filters, we show in Fig.~\ref{fig:spectrum} the spectrum of the EFIE (after curing the low-frequency breakdown to render the figure more readable), the spectrum of the filters with their logarithmically growing supports (we indicate synthetically $\mat Q^{\Sigma}$ and $\mat Q^{\Lambda}$ the two operator sums in \eqref{eq:6}. We also show the spectrum of the preconditioned symmetric blocks for blocks that clearly show why the preconditioning is working: in each logarithmically growing spectral range, the spectrum grows only of approximately a factor 2 before being rescaled again to 1 in the next range.
  The preconditioner has also been tested on a set of refined  geometries as in Fig.~\ref{fig:morphing} left. The condition numbers of the EFIE matrix for \num{1638}, \num{3210}, and \num{4785} unknowns are \emph{after} quasi-Helmholtz solution of the low-frequency breakdown \num{184}, \num{396}, and \num{553}; by further adding the action of the new Laplacian filters, the condition numbers become \num{8},  \num{5}, and \num{6} that are not far from the theoretical foreseen value of 2. 
\section*{Acknowledgment}
This work was supported by the European Research Council (ERC) through the European Union’s Horizon 2020 Research and Innovation Programme under Grant 724846 (Project 321).



%

\bibliographystyle{IEEEtran}
\bibliography{references}

\begin{thebibliography}{1}
\providecommand{\url}[1]{#1}
\csname url@samestyle\endcsname
\providecommand{\newblock}{\relax}
\providecommand{\bibinfo}[2]{#2}
\providecommand{\BIBentrySTDinterwordspacing}{\spaceskip=0pt\relax}
\providecommand{\BIBentryALTinterwordstretchfactor}{4}
\providecommand{\BIBentryALTinterwordspacing}{\spaceskip=\fontdimen2\font plus
\BIBentryALTinterwordstretchfactor\fontdimen3\font minus
  \fontdimen4\font\relax}
\providecommand{\BIBforeignlanguage}[2]{{%
\expandafter\ifx\csname l@#1\endcsname\relax
\typeout{** WARNING: IEEEtran.bst: No hyphenation pattern has been}%
\typeout{** loaded for the language `#1'. Using the pattern for}%
\typeout{** the default language instead.}%
\else
\language=\csname l@#1\endcsname
\fi
#2}}
\providecommand{\BIBdecl}{\relax}
\BIBdecl

\bibitem{8879168}
L.~Rahmouni and F.~P. Andriulli, ``{A New Preconditioner for the EFIE Based on
  Primal and Dual Graph Laplacian Spectral Filters},'' in \emph{2019
  International Conference on Electromagnetics in Advanced Applications
  (ICEAA)}, 2019, pp. 1342--1344.

\bibitem{9580445}
S.~B. Adrian, A.~Dély, D.~Consoli, A.~Merlini, and F.~P. Andriulli,
  ``Electromagnetic integral equations: Insights in conditioning and
  preconditioning,'' \emph{IEEE Open Journal of Antennas and Propagation},
  vol.~2, pp. 1143--1174, 2021.

\bibitem{9539728}
D.~Consoli, A.~Merlini, and F.~P. Andriulli, ``A fast quasi-conformal mapping
  preconditioner for electromagnetic integral equations,'' in \emph{2021
  International Conference on Electromagnetics in Advanced Applications
  (ICEAA)}, 2021, pp. 412--412.

\end{thebibliography}

\end{document}